\documentclass[prl,showpacs,twocolumn,floatfix,superscriptaddress]{revtex4}
\usepackage{times,amsmath,amsfonts,amssymb,latexsym}

\usepackage{graphicx,epsf}
\usepackage{bm}

\newcommand{\bra}[1]{\langle #1 |}

\newcommand{\ket}[1]{| #1 \rangle}

\newcommand{\be}{\begin{equation}}

\newcommand{\ee}{\end{equation}}

\newcommand{\ba}{\begin{eqnarray}}

\newcommand{\ea}{\end{eqnarray}}

\newcommand{\ignore}[1]{}

\def\CC{{\rm\kern.24em \vrule width.04em height1.46ex depth-.07ex

    \kern-.30em C}}

\def\P{{\rm I\kern-.25em P}}

\def\RR{{\rm

         \vrule width.04em height1.58ex depth-.0ex

         \kern-.04em R}}

\def\bbbc{{\mathchoice {\setbox0=\hbox{$\displaystyle\rm C$}\hbox{\hbox

to0pt{\kern0.4\wd0\vrule height0.9\ht0\hss}\box0}}

{\setbox0=\hbox{$\textstyle\rm C$}\hbox{\hbox

to0pt{\kern0.4\wd0\vrule height0.9\ht0\hss}\box0}}

{\setbox0=\hbox{$\scriptstyle\rm C$}\hbox{\hbox

to0pt{\kern0.4\wd0\vrule height0.9\ht0\hss}\box0}}

{\setbox0=\hbox{$\scriptscriptstyle\rm C$}\hbox{\hbox

to0pt{\kern0.4\wd0\vrule height0.9\ht0\hss}\box0}}}}

\def\bbbz{{\mathchoice {\hbox{$\sf\textstyle Z\kern-0.4em Z$}}

{\hbox{$\sf\textstyle Z\kern-0.4em Z$}}

{\hbox{$\sf\scriptstyle Z\kern-0.3em Z$}}

{\hbox{$\sf\scriptscriptstyle Z\kern-0.2em Z$}}}}

\begin{document}
\title{Thermal states of the Kitaev honeycomb model: a Bures metric analysis}
\author{Damian F. Abasto} \email{abasto@usc.edu}
\affiliation{Department of Physics and Astronomy, Center for Quantum Information Science\&Technology, University of Southern California, Los Angeles, CA 90089-0484, USA.}
\author{Paolo Zanardi$^{1,}$}
\affiliation{
Institute for Scientific Interchange, Viale
Settimio Severo 65, I-10133 Torino, Italy.
}

\date{September 10, 2008}

\begin{abstract}
We analyze the Bures metric over the canonical thermal states for the Kitaev honeycomb mode. In this way the effects of finite temperature on topological phase transitions can be studied. Different regions in the parameter space of the model can be clearly identified in terms of different temperature scaling behavior of the Bures metric tensor. Furthermore, we show a simple relation between the metric elements and the crossover temperature between the quasi-critical and the quasi-classical regions. These results extend the ones of  \cite{F-china,G-china} to finite temperatures. 

\end{abstract}


\maketitle
\section{Introduction}
In the last few years the use of the notion {\em fidelity} between quantum states to analyze quantum phase transitions (QPTs) has proven to be very fruitful \cite{F 0}-\cite{F 25}. The motivation comes from the fact that a major change in the ground state of a many-body system takes place near a quantum phase transition \cite{QPT}. This statement can be made quantitative by borrowing  concepts  from quantum information, in particular,  distinguishability measures between quantum states \cite{NC}. 
The simplest of such measures is provided by the amplitude of the overlap between the two states. The quantum phase transition can be detected by a drop in the fidelity between two ground states corresponding to slightly different values of the parameters defining the system. Moreover, this procedure gives rise to an associated metric tensor $g$ in the parameter space associated with the system Hamiltonian, whose divergences can signal quantum critical points.  This strategy has been successfully applied to many different systems. In particular Refs   \cite{F 14}, \cite{F-china}-\cite{AHZ} deal  with quantum phase transitions involving topological order \cite{Wen}, where a local order parameter does not exist. 

One can further generalize this fidelity approach to study how temperature affects the quantum phase transition. This is done by considering the Uhlmann fidelity \cite{Uhl} between Gibbs states corresponding to the many-body quantum system.
In \cite{F 12,Chernoff} it has been shown that different regions in the parameter space, called quasiclassical and quantum critical, can be clearly identified by  the different scaling behaviors of the resulting Bures metric tensor $g$ with temperature.

In this present work we study the effects of temperature on the topological phase transition in the Kitaev honeycomb model \cite{Kitaev}. 
We extend the results of \cite{F-china,G-china} to finite temperature by considering the metric tensor in the parameter space that comes from the Uhlmann fidelity between nearby Gibbs states. We find indeed that different regions in the parameter space can be identified according to different scalings in temperature of the metric elements, and moreover, we point out that the crossover behavior near the critical point can be identified by considering a ratio of metric elements.


Let's begin by reviewing  some of the background material supporting the idea
that near a quantum phase transition the degree of distinguishability between mixed states of a many-body Hamiltonian is enhanced. The mixed states $\rho$ we will be interested in depend on a set of parameters, e.g. tunable couplings, 
$\{\lambda_{\mu}\}$ that defines the hamiltonian of the system. 
The problem of distinguishing nearby quantum states can then be recast into a problem of estimating the value of the parameters $\{\lambda_{\mu}\}$ \cite{BC}. The estimation is made from the probability distribution of the possible results of measurements performed on the many-body states $\rho(\{\lambda_{\mu}\})$. To be concrete, let's assume that generalized measurements $\{E(\epsilon_i)\}$ are performed on the states $\rho(\{\lambda_{\mu}\})$, with possible results $\{\epsilon_i\}$. This gives rise to the probability distribution $p(\epsilon_i|\{\lambda_{\mu}\})=\mathrm{Tr}(E(\epsilon_i)\rho(\{\lambda_{\mu})\})$ for $\epsilon_i$, given the parameter $\{\lambda_{\mu}\}$.

A natural distance, called Fisher-Rao distance, can be introduced in this probability space and is given by
\begin{equation}
ds^2 = \frac{1}{4}\sum_i \frac{dp_i dp_i}{p_i},\label{W}
\end{equation}

\noindent which induces a metric $g_{\mu\nu}$ onto the parameter space $\{\lambda_{\mu}\}$ given by
\begin{equation}
g_{\mu\nu} = \frac{1}{4}\sum_i\frac{1}{p(\epsilon_i|\lambda_{\mu})}\big(\frac{\partial p(\epsilon_i|\lambda_{\mu})}{\partial \lambda_{\mu}}\big)\big(\frac{\partial p(\epsilon_i|\lambda_{\nu})}{\partial \lambda_{\nu}}\big).
\end{equation}

The Bures distance between two density matrices $\rho$ and $\rho +d\rho$ is the natural distinguishability measure that arises as the maximization of (\ref{W}) with respect to all the possible generalized measurements
\begin{equation}\label{Buresdistance}
ds^2_{\mathrm{Bures}}(\rho, \rho + d\rho)=\mathrm{max}_{\{E(\epsilon)\}}\sum_{\mu\nu}g_{\mu\nu}d\lambda_{\mu}d\lambda_{\nu}.
\end{equation}

Following Refs \cite{F 12,Chernoff}
this last expression can be expanded in the following way: $ ds^2=\sum_{\mu\nu} g^{c}_{\mu\nu}d\lambda_{\mu}d\lambda_{\nu}+g^{nc}_{\mu\nu}d\lambda_{\mu}d\lambda_{\nu},
$
with the classical and nonclassical metric elements $g^c_{\mu\nu}$ and $g^{nc}_{\mu\nu}$ given by
\begin{eqnarray}
 g^c_{\mu\nu} &=& \frac{1}{4} \sum_{i} \frac{\partial_{\mu}p_{i}\partial_{\nu}p_i}{p_{i}}\nonumber\\ 
g^{nc}_{\mu\nu}&=&\frac{1}{2} \sum_{i \neq j,\mu} \frac{(p_{i}-p_{j})^2}{p_{i}+p_{j}}|\langle i|\partial_{\mu}|j\rangle||\langle i|\partial_{\nu}|j\rangle|\label{bl},
\end{eqnarray}
where we have used the diagonal basis for $\rho$, i.e. $\rho=\sum_i p_i\ket{i}$, with $p_i$ its eigenvalues, together with the dependence of these states on the set $\{\lambda_{\mu}\}$. We see that the line element separates into a classical and a nonclassical part, the first one depending only on the probability distribution $\{p_i\}$,  while the second one depends on the set of eigenstates $\{\ket{i}\}$ themselves. The classical part is still the Fisher-Rao  distance between nearby probability distributions $p_i$ and $p_i + dp_i$, after the maximization over the generalized measurements is performed \cite{IB}. The expression for these measurement operators is known, and turns out to be a projective measurement associated to the Hermitian observable \cite{FCaves}
\begin{equation}\label{Ob}
M=\frac{1}{\sqrt{\rho}}\sqrt{\sqrt{\rho}(\rho+d\rho)\sqrt{\rho}}\frac{1}{\sqrt{\rho}}.
\end{equation}

Let's consider the expectation value of $M$ for the thermal state $\rho$:
\begin{equation}
\mathrm{Tr}(\rho M)=\mathrm{Tr}\sqrt{\sqrt{\rho}(\rho+d\rho)\sqrt{\rho}}=1-\frac{1}{2}ds^2_{\mathrm{Bures}},
\end{equation}

after expanding this expression in $d\rho$ \cite{BC}. Since the metric elements diverge at the critical point for $T=0$, or have a finite peak for $T\ne 0$, we see that in principle,  by measuring this observable  it would be possible to detect the quantum phase transition and getting an optimal estimation of the Hamiltonian parameters \cite{F 13}. Notice that $M$ depends on the (possibly unknown) state $\rho$; 
therefore the experimental realization of this measurement may require iterative strategies \cite{F 13}. 
We point out that the divergence of the metric elements across a QPT is a sufficient, but not a necesary condition \cite{F 6}.\\

In the next section we will obtain the explicit form of the Bures  metric elements for the thermal states of the Kitaev honeycomb model.
\section{Bures metric and the Kitaev honeycomb model}
The Kitaev honeycomb model \cite{Kitaev} is one of the most important examples of a solvable model in two dimensions that exhibits topological phases, with non-abelian anyons. We will study the effects of temperature over its critical lines, by considering the Bures metric for nearby thermal states. We will first briefly review the model and its diagonalization and then compute and analyze the associated Bures metric.

\subsection{The honeycomb model}
The model consists of spins 1/2 placed at the sites of a honeycomb lattice, with nearest neighbor interactions. There are three types of links at each site, called x, y, or z, depending on their direction in the lattice. The Hamiltonian of the model is given by
\begin{equation}
H=-J_x\sum_{x-\mathrm{links}}\sigma^{x}_{j}\sigma^{x}_{k}-J_y\sum_{y-\mathrm{links}}\sigma^{y}_{j}\sigma^{y}_{k}-J_z\sum_{z-\mathrm{links}}\sigma^{z}_{j}\sigma^{z}_{k}\label{Kitaev},
\end{equation}
where the ends of the corresponding links are labeled by $j,k$, and $J_x$, $J_y$ and $J_z$ are the corresponding model parameters. The phase diagram can be separated basically into two regions of the parameter space, given by the following inequalities\begin{equation}
|J_z|\le |J_x| + |J_y|, \; |J_y|\le |J_z| + |J_x|, \; |J_x|\le |J_y|+|J_z|\label{regions},
\end{equation}
with the equal signs signaling the lines of quantum criticality. 

The region outside these inequalities is gapped and contains abelian anyons, while the region inside is gapless and gives rise to non-abelian anyons in the presence of a magnetic field. 
To diagonalize this model, we replace the spin 1/2  at every site $i$ by four Majorana fermions, by writing each Pauli operator as $\sigma^{a}_{j}=ib^{a}_jc_j$, with $b^a$ and $c$ Majorana fermions, and $a=x, y, z$. These Majorana operators act on a 4-dimensional Fock space. This enlarged space has redundant degrees of freedom which can be projected onto a physical 2-dimensional Hilbert space $\mathcal{L}$ suitable for a spin 1/2  by imposing the condition: $\ket{\psi}\in\mathcal{L}\iff D\ket{\psi}=\ket{\psi}$, where $D$ is a projector given by $D=b^xb^yb^zc$. Within this subspace, the Hamiltonian (\ref{Kitaev}) can be written as
\begin{equation}
H=\frac{i}{2}\sum_{j,k}J_{a_{jk}}\hat{u}_{jk}c_jc_k,
\end{equation}
where the index $a_{jk}$ is x, y, or z depending on the direction of the link connecting sites $j$ and $k$, and the operators $\hat{u}_{jk}=ib_j^{a_{jk}}b_k^{a_{jk}}$ are such that $[\hat{u}_{jk},H]=0$, $[\hat{u}_{jk}, \hat{u}_{pq}]=0$ and $\hat{u}^2_{jk}=1$. The entire Hilbert space can be decomposed into eigenspaces of the operators $\hat{u}_{jk}$, indexed by the eigenvalues $u_{jk}=\pm1$. It can be proven that the configuration of the $u_{jk}$ that minimizes the ground state energy is given by the vortex-free configuration, that is, $u_{jk}=1$ for all links $(j,k)$. Given the translational symmetry of this configuration,  one can perform a Fourier transform, obtaining the Hamiltonian in momentum space within a unit cell
\begin{equation}
H=\frac{1}{2}\sum_{\mathbf{p},\lambda,\gamma} iA_{\lambda\gamma}(\mathbf{p})c^{\dagger}_{\lambda}(\mathbf{p})c_{\gamma}(\mathbf{p})\label{H},
\end{equation}
where $\mathbf{p}=(p_x,p_y)$, $c_{\lambda}(\mathbf{p})=\frac{1}{\sqrt{N}}\sum_{\mathbf{r}}e^{-i\mathbf{p}\cdot \mathbf{r}}c_{\lambda}(\mathbf{r})$ and $A(\mathbf{p})$ is a $2\mathrm{x}2$ matrix given by
\begin{equation}
iA(\mathbf{p})=\left(\begin{array}{c c} 0 & if(\mathbf{p}) \\ -i f(\mathbf{p})^* & 0
\end{array}\right),
\end{equation}
with $f(\mathbf{p}) = \epsilon(\mathbf{p}) + i\Delta(\mathbf{p})$, $\epsilon(\mathbf{p})=2(J_x\cos{p_x}+J_y\cos{p_y}+J_z)$ and $\Delta(\mathbf{p})=2(J_x\sin{p_x}+J_y\sin{p_y})$.

In these equations $\mathbf{r}$ is the position of the unit cell, while $\lambda,\gamma$ are the indices for the sites inside each cell. $N$ is the number of sites in the lattice, and we choose $N=2L^2$, with $L$ odd. The momenta take the values $p_{x(y)}=\frac{2n\pi}{L},n=-\frac{L-1}{2},\cdots,\frac{L-1}{2}$. 

We can further put the Hamiltonian (\ref{H}) in a quasi-free fermion form by introducing the fermionic operators \cite{Pachos}
\begin{eqnarray}
b(\mathbf{p})&=&\frac{1}{\sqrt{2}}\big(c_1(\mathbf{p})+ie^{i\theta}c_2(\mathbf{p})\big)\\
b^{\dagger}(\mathbf{p})&=&\frac{1}{\sqrt{2}}\big(c_1^{\dagger}(\mathbf{p})-ie^{-i\theta}c_2^{\dagger}(\mathbf{p})\big)\label{b},
\end{eqnarray}
with $\theta=\mathrm{arg}(f(\mathbf{p}))$. Then the Hamiltonian takes the following diagonal form:
\begin{equation}
H=\sum_{\mathbf{p}}\Lambda(\mathbf{p})\Big[b^{\dagger}(\mathbf{p})b(\mathbf{p})-\frac{1}{2}\Big],
\end{equation}
where $\Lambda(\mathbf{p})=|f(\mathbf{p})|=\sqrt{\epsilon(\mathbf{p})^2+\Delta(\mathbf{p})^2}$ represents the quasiparticle excitation energies. 

The ground state is given by $\ket{gs}=\prod_{\mathbf{p}}\big(c_1^{\dagger}(\mathbf{p})+ie^{-i\theta}c_2^{\dagger}(\mathbf{p})\big)\ket{0}$, with $\ket{0}$ the  vacuum state of the Majorana operators $c_{\lambda}(\mathbf{p})$. The excited states are created by applying $b^{\dagger}(\mathbf{p})$ onto the vacuum state.
\subsection{Bures metric over thermal states}

We are interested in characterizing the effects of finite temperature on the quantum phase transitions of this model. We consider then the Kitaev model in thermal equilibrium with a heat bath at temperature $T$, and obtain the expression for the Bures metric (\ref{bl}) induced onto the 4-dimensional parameter space of the Kitaev model $\{\beta,J_a\}$, with $a=x,y,z$ and $\beta=1/T$.  In this situation, its states will be described by $\rho(T,\{J_a\})=Z^{-1}e^{-\beta H}=Z^{-1}\sum_ie^{\beta E_i}\ket{i}\bra{i}$, with $Z=\mathrm{Tr}e^{-\beta H}$, $E_i$ and $\ket{i}$ the eigenvalues and eigenvectors of the Hamiltonian. Using the equations ($\ref{bl}$), the expression for the fermionic operators ($\ref{b}$), together with the general results in \cite{F 12}, we obtain the following classical metric elements in the thermodynamic limit

\begin{eqnarray}\label{class}
g^{c}_{\beta\beta}&=&\frac{1}{32\pi^2}\int_{-\pi}^{\pi}dp_xdp_y\frac{1}{\cosh(\Lambda\beta)+1}\Lambda^2\nonumber\\
g^{c}_{\beta J_{a}}&=&\frac{\beta}{32\pi^2}\int_{-\pi}^{\pi}dp_xdp_y\frac{1}{\cosh(\Lambda\beta)+1} \Omega_{a}\nonumber\\
g^{c}_{\beta J_{z}}&=&\frac{\beta}{32\pi^2}\int_{-\pi}^{\pi}dp_xdp_y\frac{1}{\cosh(\Lambda\beta)+1}2\epsilon\nonumber\\
g^{c}_{J_{a} J_{a}}&=&\frac{\beta^2}{32\pi^2}\int_{-\pi}^{\pi}dp_xdp_y\frac{1}{\cosh(\Lambda\beta)+1}\frac{\Omega^2_{{a}}}{\Lambda^2}\nonumber\\
g^{c}_{J_zJ_z}&=&\frac{\beta^2}{32\pi^2}\int_{-\pi}^{\pi}dp_xdp_y\frac{1}{\cosh(\Lambda\beta)+1}\Big(\frac{2\epsilon}{\Lambda}\Big)^2\nonumber\\
g^{c}_{J_x J_{y}}&=&\frac{\beta^2}{32\pi^2}\int_{-\pi}^{\pi}dp_xdp_y\frac{1}{\cosh(\Lambda\beta)+1}\frac{\Omega_x\Omega_y}{\Lambda^2}\nonumber\\
g^{c}_{J_{a} J_z}&=&\frac{\beta^2}{32\pi^2}\int_{-\pi}^{\pi}dp_xdp_y\frac{1}{\cosh(\Lambda\beta)+1} \Omega_{a}\frac{2\epsilon}{\Lambda^2},\nonumber\\
\end{eqnarray}
where $a\in \{x,y\}$, $\Omega_x=2(\cos(p_x)\epsilon(\mathbf{p})+\sin(p_x)\Delta(\mathbf{p}))$ and $\Omega_y=2(\cos(p_y)\epsilon(\mathbf{p})+\sin(p_y)\Delta(\mathbf{p}))$.

The quantum or nonclassical metric elements are
\begin{eqnarray}\label{quant}
g^{nc}_{J_aJ_b}&=&\frac{1}{32\pi^2}\int_{-\pi}^{\pi}dp_xdp_y\frac{\cosh(\Lambda\beta)-1}{\cosh(\Lambda\beta)+1}\frac{\Theta_a\Theta_b}{\Lambda^4}\nonumber\\
\end{eqnarray}
where $a,b \in \{x,y,z\}$, $\Theta_x=4(J_z\sin(p_x)+J_y\sin(p_x-p_y))$, $\Theta_y=-4(J_x\sin(p_x-p_y)-J_z\sin(p_y))$ and $\Theta_z=-2\Delta(\mathbf{p})$.

\begin{figure}[htp]
\centering
\includegraphics[scale=0.3]{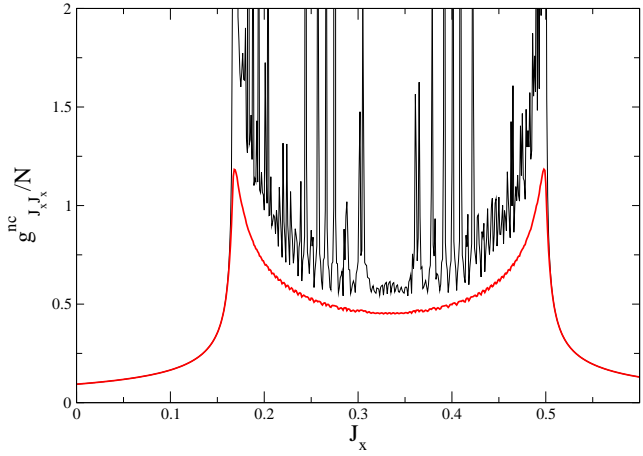}
\caption{(Color online). Plot of $g^{nc}_{J_zJ_z}/N$ for a finite size of $N = 2L^2$, $L=101$, as a function of $J_x=2/3-J_y$. The black and red curves correspond to $T=0$ and $T=0.01$, respectively. As soon as temperature is finite, the peaks (due to finite size effects) are suppressed.}\label{fig:metricplot}
\end{figure}

In references \cite{FC,GC} the authors showed that a related quantity, the fidelity susceptibility, could characterize the quantum phase transition between these two phases at zero temperature, by exhibiting a divergence at the critical lines (\ref{regions}). Indeed, similar results can be obtained by taking the limit $T\rightarrow 0$ in (\ref{class}) and (\ref{quant}). Furthermore, since in this limit the metric elements are inversely proportional to the energy gap \cite{F 5}, they exhibit a series of peaks in the gapless region, at values of the momenta $\mathbf{p_0}$ for which $\Lambda=0$ \cite{GC}. In the case of finite temperature the metric elements don't exhibit divergences but finite peaks at the critical points, while the peaks in the gapless region are smoothed out, since all the momenta $\mathbf{p_0}$ are occupied with a probability given by the Boltzmann  distribution. Figure \ref{fig:metricplot} shows a plot of the metric element $g^{nc}_{J_zJ_z}$, for a finite size of $N = 2L^2$, $L=101$ \cite{G-china}, for $T=0$ and $T=0.01$. As can be seen, as soon as the temperature is finite, the peaks caused by finite size effects are smoothed out. \\
\\

\subsection{Temperature analysis of the metric elements}

In the first part of this section we will characterize two different regimes in the parameter space $\{\beta, J_a\}$, namely, the quasi-classical and quantum critical, in terms of the different scaling behavior of the metric elements with temperature. The quasi-classical region is located away from criticality, for the gapped phases outside the region ($\ref{regions}$), and the range of temperatures are such that $\beta\Delta\gg1$, with $\Delta$ the fermionic gap of the system, given by $\Delta=2(J_z-J_x-J_y)$ for $|J_z|\ge |J_x| + |J_y|$, and similarly for the other two gapped regions. The quantum critical region is located at the lines of criticality that separate the gapped and the gapless phases, at finite temperature. For this last case, we will analyze the scaling behavior of the metric elements for $\Delta=0$, and approaching the quantum critical point by taking $T \rightarrow 0$.

Let's proceed first with the scaling of the classical metric elements in the quasi-classical region. We will focus mainly on the region $|J_z|\ge |J_x| + |J_y|$, but the analysis is very similar for the other two gapped phases. Since $\beta\Lambda(\mathbf{p})>\beta\Delta\gg1$, we can do the following approximation: $\frac{1}{\cosh(\beta\Lambda)+1}\approx2e^{-\beta\Lambda(\mathbf{p})}$. The exponential $e^{-\beta\Lambda(\mathbf{p})}$ has a sharp peak centered around the minima of $\Lambda(\mathbf{p})$, which is given by $\mathbf{p}=(\pm \pi, \pm \pi)$. The asymptotic behavior of the metric elements with temperature can be obtained by performing a saddle point approximation, by expanding $\Lambda(\mathbf{p})$ to second order, the rest of the integrand up to the first nonzero term around the minima points, and extending the limits of integration in $p$ from $0$ to $\infty$. This results in the following scaling behavior of the classical metric elements

\begin{equation}
g^{c}_{ab}(\beta\Delta\gg1)\approx T^{\alpha}e^{-\Delta/T},
\end{equation}
 with $\alpha =1$ for $g^{c}_{\beta\beta}$, $\alpha=0$ for $a=\beta$, $b=J_x,J_y, J_z$ and $\alpha=-1$ for $a,b = J_x,J_y, J_z$.

A similar analysis can be performed for the nonclassical metric elements, with the result
\begin{equation}
g^{nc}(\beta\Delta\gg1)\approx g^{nc}(T=0)+f(J_x,J_y,J_z)T^2e^{-\Delta/T},
\end{equation}
with the same exponent, $T^2$, for all these metric elements, with a nonuniversal function $f(J_x,J_y,J_z)$. 

We focus now on the quantum critical region. We will perform the scaling analysis with temperature inside and along the lines of criticality that separate the gapped from the gapless phase, and then take the limit $T \rightarrow 0$. In this limit, all classical metric elements vanish, so we are left to analyze the nonclassical ones. 

Inside the gapless region (\ref{regions}), the scaling with temperature is dominated by the divergence of $\Lambda^{-4}$ at its zeroes, given by $p_x^{0}=\pi \pm \arccos\Big(\frac{J_x^2+J_z^2-J_y^2}{2J_xJ_z}\Big)$ and $p_y^{0}=\pi \pm \arccos\Big(\frac{J_y^2+J_z^2-J_x^2}{2J_yJ_z}\Big)$, around which $\Lambda\approx \alpha(p_x-p_x^{0}) + \beta(p_y-p_y^{0})$, with $\alpha$ and $\beta$ constants. The asymptotic behavior of the integrals can be well captured if we shift the limits of integration from $(-\pi, \pi)$ to $(0, 2\pi)$, and focus on the region around $\mathbf{p}=(p_x^{0}, p_y^{0})$. The factor $\frac{\cosh(\Lambda\beta)-1}{\cosh(\Lambda\beta)+1}$ presents a sharp drop at $(p_x^{0}, p_y^{0})$ that gets steeper the smaller the temperature T is, and away from this point is equal to $1$. For the purposes of obtaining the scaling of the metric elements, it can be very well represented by the following piecewise function

\begin{displaymath}
\frac{\cosh(\Lambda\beta)-1}{\cosh(\Lambda\beta)+1}\approx \left\{ \begin{array}{ll}
p^2/T^2 & \textrm{for $p \le T$}\\
1 & \textrm{otherwise,}
\end{array} \right.
\end{displaymath}
where we have taken polar coordinates $(p,\theta)$, centered around $(p_x^{0}, p_y^{0})$. Performing a Laurent expansion of $\frac{\Theta_a\Theta_b}{\Lambda^4}$, it can be seen that the nonclassical metric elements diverge logarithmically inside the gapless region
\begin{equation}
g^{nc}_{ab}(\Delta=0, T\rightarrow 0)\approx \ln(T).
\end{equation}

 This logarithmic divergence is related to the 2D nature of the model, and is different from power law behaviors that have been reported for the Ising \cite{F 12} and XY chains \cite{Chernoff} at finite temperature.

Along the phase boundaries (\ref{regions}), there is always a direction in momentum space along which the quasiparticle energy $\Lambda(\mathbf{p})$ is no longer linear but quadratic in momentum. For example, along the path $J_x=J_y=\frac{1-J_z}{2}$, $\Lambda(\mathbf{p})\approx (p-\pi)^2$ for $p_x=-p_y$. This quadratic dispersion dominates the scaling of the nonclassical metric elements with temperature, so that if one restricts the double integrals in (\ref{quant}) to a single integral along these paths in momentum space for which the dispersion becomes quadratic, one obtains that 
\begin{equation}
g^{nc}_{ab}(\Delta=0, T\rightarrow 0)\approx T^{-1/2},
\end{equation}
 
\noindent at the boundaries between the phases, dominating the logarithmic divergence. We have checked this argument numerically and confirmed that indeed the nonclassical metric elements diverge as $T^{-1/2}$ for $T\rightarrow 0$.

We conclude that different regimes in the parameter space $\{\beta, J_a\}$ can indeed be identified by the distinct scaling in temperature of the Bures metric elements. We stress again that the nonclassical metric elements display a logarithmic divergence with temperature.

\section{Quantum versus Classical contributions}

\begin{figure}[htp]
\centering
\includegraphics[scale=0.26]{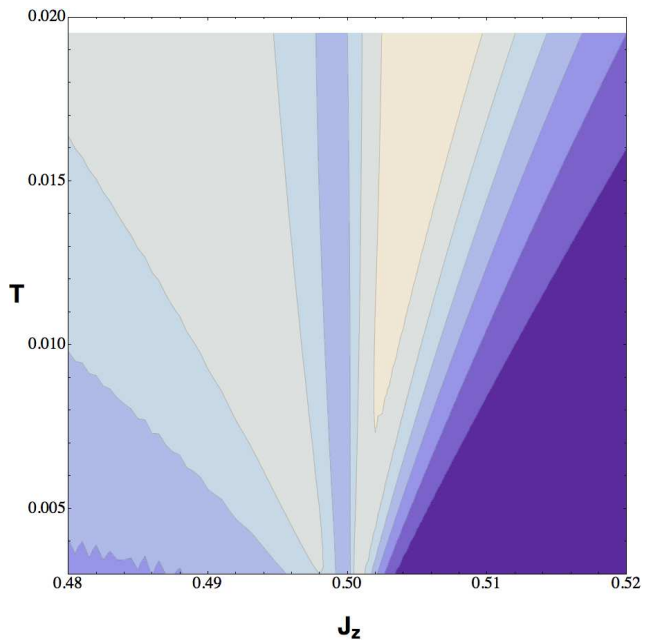}
\caption{(Color online). Contour plot of the ratio $g_{J_zJ_z}/g_{J_zJ_z}^{nc}$ as a function of $J_z$ and temperature $T$, along the trajectory $J_x=J_y=\frac{1-Jz}{2}$. There is an enhancement in discrimination due to the nonclassical metric element in the quantum critical region as well as away from it, for small temperatures. The color scale goes from 0.1 (purple), to 0.6 (white).}\label{fig:contour}
\end{figure}

We now remind the reader about the interpretation of the Bures metric elements. As we have seen, the Bures distinguishability metric depends on two parts, one classical and one nonclassical. The classical metric elements characterize the distinguishability between nearby density states due to changes in the mixing probabilities $p_i$ of the density matrix $\rho$, while the nonclassical metric elements characterize how the distniguishability is enhanced by the changes in the quantum states $\ket{i}$ themselves. We can therefore consider the ratio between the classical and corresponding nonclassical metric elements as a figure of merit to determine the relative impact on the overall distinguishability due to each classical and quantum contribution. Figure \ref{fig:contour} shows a contour plot for the ratio $g_{J_zJ_z}^c/g_{J_zJ_z}^{nc}$ in the plane $(J_z, T)$, along the trajectory $J_x=J_y=\frac{1-Jz}{2}$, for $0.48 < J_z < 0.52$. In the regions away from the quantum critical point $J_z=0.5$ and small temperatures and at the quantum critical region above $J_z=0.5$ we have that the nonclassical metric element has a predominant contribution to the Bures line element, while for regions in between the classical metric element prevails. 

The contour plot seems to capture the crossover behavior, with a crossover at $T\sim | J_z - 0.5 |^{z\nu}$, with $z = \nu = 1$ \cite{Kitaev,F-china,G-china}. Indeed, using the expressions for the classical and nonclassical metric elements developed in \cite{F 12}, one can prove that near a quantum critical point $\frac{g^c}{g^{nc}} \sim \big(\frac{\Delta}{T}\big)^2$, with $\Delta$ being the gap. Since near quantum criticality $\Delta \sim |g - g_c |^{z\nu}$ \cite{QPT}, with $g$ the coupling constant driving the transition at $g_c$, we have that a contour plot would reveal the crossover at $T\sim | g - g_c |^{z\nu}$.


\section{Conclusions}
In this work we have analyzed the canonical thermal states of the Kitaev honeycomb model by using the Bures metric tensor. We have shown that  metric elements can be used to distinguish two regions in the parameter space $\{T, J_a\}$, namely the quasi-classical and quantum critical regions, by different temperature scaling of the metric elements. A novel logarithmic divergence with vanishing temperature has been  obtained for the nonclassical part of the metric elements. 

Furthermore, the ratio between classical and quantum parts of metric elements has been used as a figure of merit to compare the classical and quantum contributions to the overall distinguishability of nearby states. It is interesting to point out that similar behavior for such ratios between metric elements can be seen for the Ising and XY models at finite temperature. It is tempting to compare this behavior with the usual crossover phenomena  that separate regions in which the fluctuations of order parameters are thermal or quantum. Nevertheless, this connection is blurred for the present model, since there is no local order parameter for the Kitaev honeycomb model. An interesting open question for future investigations is to see whether information metric
techniques similar to those exploited in these paper can help in studying quantum phase transitions involving different types of topological order see e.g., \cite{nonabany}.

\section{acknowledgements}
We thank N. Toby Jacobson for a careful reading of the manuscript. We acknowledge financial support 
by the National Science Foundation under grant DMR- 
0804914, and 0803304.

\end{document}